\documentclass[prb, longbibliography, twocolumn]{revtex4-2}
\usepackage{bm}
\usepackage{graphicx}
\usepackage{amsmath}
\usepackage{amssymb} 
\usepackage[utf8]{inputenc} 
\usepackage[T1]{fontenc}
\usepackage{color}
\usepackage{xcolor}
\usepackage{upgreek}
\usepackage{subfigure}
\usepackage[unicode=true,colorlinks=true,citecolor=blue]{hyperref}
\usepackage{lipsum}
\usepackage{epsfig}
\usepackage{wrapfig}
\usepackage[normalem]{ulem}
\usepackage{units}
\usepackage{cancel}
\usepackage{float}
\usepackage{lipsum}
\usepackage{placeins}
\usepackage{ocgx}

\newcommand{\nix}[1]{}

\renewcommand{\phi}{\varphi}

\renewcommand{\i}{\mathrm i}

\newcommand{\eps}{\varepsilon}

\newcommand{\beq}{\begin{equation}}
	\newcommand{\eeq}{\end{equation}}
\newcommand\beqa{\begin{eqnarray}}
	\newcommand\eeqa{\end{eqnarray}}
\newcommand\ba{\begin{array}}
	\newcommand\ea{\end{array}}

\begin{document}
	
	\title{Spin separation of Dirac electrons by twisted and vector optical beams}

	\author{A. A. Gunyaga}
	\affiliation{Ioffe Institute, 194021 St. Petersburg, Russia}
	
	\author{M. V. Durnev} 
	\affiliation{Ioffe Institute, 194021 St. Petersburg, Russia}
	
	\author{S. A. Tarasenko} 
	\affiliation{Ioffe Institute, 194021 St. Petersburg, Russia}
	
	\begin{abstract}
	We demonstrate that twisted and vector optical beams can spatially separate two-dimensional electrons with opposite spin projections. Spin accumulation arises from diverging  spin currents induced by structured light together with spin relaxation and diffusion. We derive analytical expressions for the resulting spin density in the quasi-local and diffusion-controlled regimes and numerically calculate the spin textures for the parameters relevant to transition metal dichalcogenide monolayers. Our results suggest a mechanism by which the twist and vector pattern of optical beams can be imprinted onto the electron spin degree of freedom.	

	\end{abstract}
	
	\maketitle
	
	
	\textit{Introduction.} There is a growing interest in incorporating twisted light carrying orbital angular momentum (OAM) in solid state spintronics~\cite{Quinteiro-Rosen:2022, Fujita:2017, Watzel:2018, Sirenko:2019, Solyanik-Gorgone:2019, Waetzel2020, Sirenko:2021, Gunyaga:2026}.
	However, the spin selective manipulation of electrons by the phase structure of electromagnetic field in the optical beam is still challenging~\cite{Cygorek:2015,Solyanik-Gorgone:2019,Grass:2022}.
	The fundamental obstacle is that electron-photon interaction in semiconductors is primarily local, far below
         the scale of light wavelength where the parameters of structured light, such as polarization or phase, vary.
	Corrections beyond the local electro-dipole approximation, such as magneto-dipole or electro-quadrupole transitions are typically tiny (of the order of $(k \lambda)^{-2}$, where $k$ is the electron wave vector and
	$\lambda$ is the light wavelength) and give very low efficiency of the direct transfer of the photon OAM to the electron spin system~\cite{Babiker:2002, Scholz-Marggraf:2014, Schmiegelow:2016, Giammanco:2017, Kiselev:2024}.
	
	Here, we propose and study a different mechanism of spin manipulation by structured light. In this mechanism, structured light induces textured spin currents, 
	controlled by the field spatial structure, which in turn give rise to spin separation with local spin density. Although this mechanism is indirect, it can generate spin more efficiently than the direct mechanism by a factor of $k l_e$, where $l_e$ is the electron mean free path.
	We consider the spin separation for the class of two-dimensional (2D) Dirac materials excited by twisted beams 
	carrying OAM and vector beams with polarization vorticity. In both cases, the electromagnetic field is locally linearly polarized throughout the beam cross section, while the beams possess a characteristic twist or vorticity. This allows us to exclude the standard optical spin orientation by 
	circularly polarized light~\cite{OO_book} and focus on the effects associated with light twist or polarization vorticity. We also neglect possible inversion-symmetry breaking and the associated spin-orbit coupling, which can lead to uniform spin currents~\cite{Ganichev2006,Ivchenko2008} and uniform optical orientation by linearly polarized light~\cite{Tarasenko2005,Gorelov2011},  unrelated to the nontrivial spatial structure of the beam.

	\textit{Microscopic theory.} We consider two classes of paraxial optical beams: (i) twisted beams carrying orbital angular momentum (OAM)
	and (ii) vector beams with polarization vortices. Although in both cases the electromagnetic field is locally linearly polarized throughout the cross section, the    
	beams have a certain characteristic of twist or vorticity as shown below.
	
	(i) The first class are the twisted Bessel beams with the complex amplitude of the electric field in the cross section
	\begin{equation}\label{El}
		\bm E (r, \varphi) =  J_l(qr) e^{\i l\varphi} \bm E_0 \,,
	\end{equation}
	where $r = \sqrt{x^2 + y^2}$ and $\varphi = \arctan (y/x)$ are the polar coordinates, $J_l$ is the Bessel function, $l$ (integer number) is the OAM projection, 
	$q$ is the in-plane wave vector, and  $\bm E_0 = E_0 (\cos\varphi_0,\sin\varphi_0)$ is the vector describing 
	the field amplitude and linear polarization in the $(x,y)$ frame. The length $L = 2\pi/q$ determines the radii of bright and dark intensity rings in the beam cross section.
	
	(ii) The second class are the vector beams with the electric field amplitude
	\begin{equation}\label{En}
		\bm E (r,\varphi) =  J_n(qr) E_0 \, \bm e_n(\varphi)  \,,
	\end{equation}
	where $\bm e_n(\varphi) = \bigl[\cos(n\varphi + \varphi_0),\,\sin(n\varphi+\varphi_0)\bigr]$ is the unit polarization vector, which rotates with the polar angle.
	The 
	(integer) winding number $n$ describes the net number of full turns that the vector $\bm e_n$ makes as the polar angle $\varphi$ varies from $0$ to $2\pi$, 
	the sign of $n$ shows the rotation direction.
	The angle $\varphi_0$ defines the field polarization at $\varphi = 0$. We note that the vector beams of the form~\eqref{En} can be created as 
	the superpositions of two circularly polarized twisted beams with the opposite OAM and the opposite helicity~\cite{Nesterov2000,Maurer:2007}.

	Mechanism of the spin separation is the following.
	The absorption of structured light with spatially varying phase (as in the case of twisted beams) or spatially varying linear polarization (as in the case of vector beams)
	in 2D Dirac materials generates local spin currents $\bm j_s = \bm i_{+1/2} - \bm i_{-1/2}$, where $\bm i_{\pm 1/2}$ are the electron fluxes in the $s = \pm 1/2$ 
	spin bands~\cite{Gunyaga:2026}. 
	Interplay between spin current generation, spin diffusion, and spin relaxation results in the emergence of a
	steady-state spin density $s(\bm r)$, which is found from the diffusion-relaxation equation
	\begin{equation}\label{diffusion_equation}
		D_s \Delta s(\bm r) - \frac{s(\bm r)}{\tau_s} = \mathrm{div} \bm j_{s} (\bm r) \,,
	\end{equation}
	where $\Delta = \partial_x^2 + \partial_y^2$ is the 2D Laplace operator, $D_{s}$ is the spin diffusion coefficient, and $\tau_s$ is the spin relaxation time. 
	Solution of Eq.~\eqref{diffusion_equation} for the general case is obtained by the Green function method and reads
	\begin{equation}\label{general solution}
		s(\bm r) = -\frac{1}{2\pi D_s}\int\!\mathrm{d}\bm r' K_0 \left( \frac{|\bm r-\bm r'|}{l_s} \right)\,\mathrm{div}\bm j_{s}(\bm r') \,,
	\end{equation}
	where $K_0(x)$ is the zero-order Macdonald function (modified Bessel function of the second kind) and $l_s = \sqrt{\tau_sD_s}$ is the spin diffusion length.
	
	The spatial scales of the integrands $K_0$ and $\mathrm{div}\bm j_{s}$ in Eq.~\eqref{general solution} are determined by $l_s$ and $L = 2\pi /q$, respectively.
	At $l_s \ll L$, one obtains  
	\begin{equation}\label{s_local}
		s^{\mathrm{(loc)}}(\bm r) = - \tau_s\,\mathrm{div}\bm j_{s}(\bm r) \,.
	\end{equation}
	It corresponds to the quasi-local approximation with negligible spin diffusion.
	In the opposite regime, when $l_s\gg L$, Eq.~\eqref{general solution} yields 
	\begin{equation}\label{s_diff}
		s^{\mathrm{(diff)}}(\bm r) = \frac{1}{2\pi D_s}\int\!\mathrm{d}\bm r'\,\mathrm{ln}\bigl(|\bm r-\bm r'|\bigr) \, \mathrm{div}\bm j^{s}(\bm r').
	\end{equation}
	This equation corresponds to the case when spin relaxation is neglected and, hence, the generation of the spin density is balanced by spin diffusion. 
	In that case, Eq.~\eqref{diffusion_equation} takes the form of the 2D Poisson equation.
	
	To proceed further we introduce equations for the spin currents induced by structured light in 2D Dirac systems following Ref.~\cite{Gunyaga:2026}. 
	The general phenomenological expression for $\bm j_s$ reads
	\begin{align}
		\label{js}
		\bm j_{s} =\ &\mathcal W_1\bm n\times \nabla S_0 + \mathcal W_2\bigl(\bm n\times \tilde\nabla S_1 - \tilde\nabla S_2\bigr) + \mathcal W_3\nabla S_3\nonumber\\
		&+ \mathcal W_4\,\bm n\times\mathrm{Im}\left[E_x^*\nabla E_x + E_y^*\nabla E_y\right]\nonumber\\
		&+ \mathcal W_5\,\bm n\times\mathrm{Im}\left[\bm E^*(\nabla\cdot\bm E) - \bm E^*\times(\nabla\times \bm E)\right]\nonumber\\
		&+ \mathcal W_6\,\bm n\times\mathrm{Re}\left[\bm E^*(\nabla\cdot\bm E) - (\bm E^*\cdot\nabla)\bm E\right] \,.
	\end{align}
	Here, $\mathcal W_j$ are real parameters, $S_0 = |E_x|^2+|E_y|^2$, $S_1 = |E_x|^2 - |E_y|^2$, $S_2 = E_xE_y^* + E_x^*E_y$, $S_3 = \i(E_xE_y^* - E_x^*E_y)$ are the local Stokes parameters of the incident radiation in the paraxial approximation, $\nabla = (\partial_x, \partial_y, 0)$, $\tilde\nabla = (\partial_x,-\partial_y,0)$, and $\bm n = (0,0,1)$.
	For electromagnetic fields given by Eq.~\eqref{El} or Eq.~\eqref{En}, the terms proportional to $\mathcal W_1$, $\mathcal W_3$, and $\mathcal W_6$ 
	do not contribute to $\mathrm{div}\bm j_{s}$ and, hence, do not contribute to the spin separation. The relevant parameters for 2D Dirac systems are
	\begin{eqnarray}
	\label{Wj}
		&\mathcal W_2\ &= \dfrac{e^2\tau_1 a^2}{32\delta^2w^5}(w^2-1)\left[1 + \frac{\theta\tau_2\delta}{\hbar}(w^2-1)\right] , \nonumber \\
		&\mathcal W_4\ &= \dfrac{\theta e^2\tau_1 a^2}{32\delta^2 w^6}\bigl[(w^2+1)^2 + (w^4-1)(\mathrm{ln}\,\tau_1)'\bigr] , \nonumber \\
		&\mathcal W_5\ &= \dfrac{\theta e^2\tau_1 a^2}{64\delta^2 w^6}(w^2-1)^2\bigl[1 - (\mathrm{ln}\, \tau_1)'\bigr] ,
	\end{eqnarray}
	where $e$ is the electron charge, $\tau_1$ and $\tau_2$ are the relaxation times of the first and second angular harmonics of the electron distribution function, 
	$2\delta$ is the band gap, $a$ is the interband velocity determining the effective mass $m^* = \delta/a^2$ at the band extrema, 
	$w = \hbar \omega /(2\delta)$, $(\mathrm{ln}\, \tau_1)' = \eps (d \tau_1/d\eps)/\tau_1$, $\eps = \hbar\omega/2 - \delta$ is the kinetic energy of a 
	photoelectron, 
	and $\theta$ is the spin Hall angle. The parameters $\mathcal W_{j}$ correspond to $W_{j} + (\theta/e) Q_{j}$ 
	from Tab.~1 in Ref.~\cite{Gunyaga:2026}.

	\begin{figure*}[t]
		\includegraphics[width=0.95\linewidth]{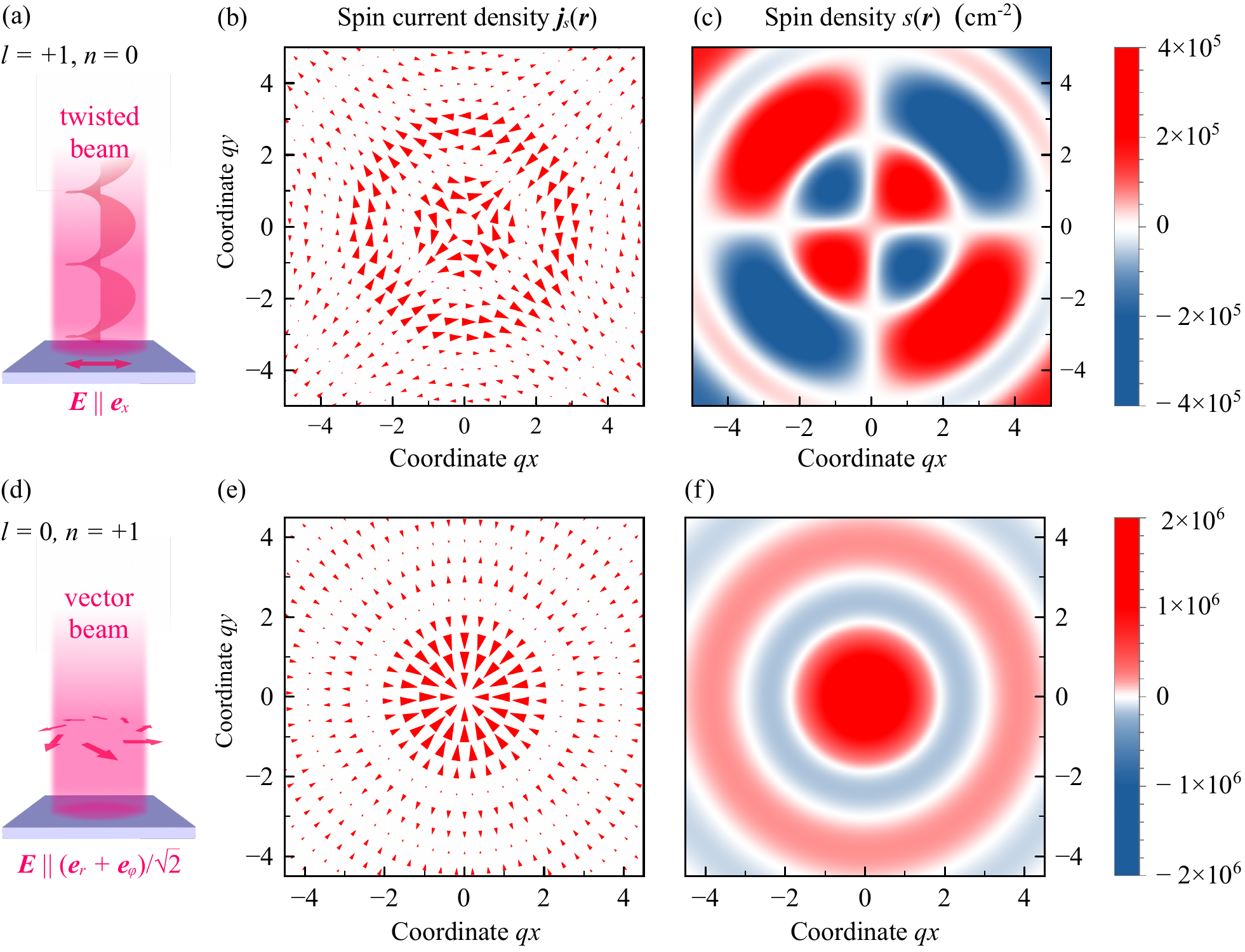}
		\caption{Optical generation of spin currents and spin polarization of electrons by linearly polarized twisted and vector beams. 
			(a) and (d)  Sketches of optical beams with orbital angular momentum $l$ and winding number $n$. 
			(b) and (e)  Spatial distributions of spin currents, (c) and (f) Spatial distributions of emerging spin densities. 
			Figures are calculated for parameters relevant to $n$-type TMDC layers: the band gap $2 \delta = 2\text{ eV}$, the interband velocity $a = c/300$, the relaxation times $\tau_1 = \tau_2 = 1\text{ ps}$, the spin diffusion coefficient $D_s = 15\text{ cm}^2/\text{s}$, the spin relaxation time $\tau_s = 10~\text{ns}$,	
			the spin Hall angle $\theta = 10^{-2}\,$rad, the radiation intensity  $I = 1\text{ kW/cm}^2$, the photon energy $\hbar\omega/2\delta = 1.1$, 
			and the in-plane wave vector $q = 0.1(\omega/c)$. Only contributions to spin currents with $\mathrm{div}\bm j_{s} \neq 0$, which give rise to spin separation, are shown.}
		\label{Figure1}
	\end{figure*}

	\textit{Results.} The details of numerical calculations of the spin density $s (\bm r)$ following Eq.~\eqref{general solution} as well as analytical equations obtained in the quasi-local 
	and diffusion-controlled approximations are given in Appendix.
	Here, instead, we focus on the results. 
	
	Figure~\ref{Figure1} shows the spatial distributions of the spin current $\bm j_s(\bm r)$ and the electron spin density $s(\bm r)$ emerging in a 2D Dirac system at
	irradiation by structured optical beams. Figures~\ref{Figure1}(a)-(c) present the results for the twisted beam with the OAM projection $l=1$ and 
	the global linear polarization along $x$. Figures~\ref{Figure1}(d)-(f) show the spin current and spin density distributions induced by the linearly polarized vector beam with the winding number $n=1$ 
	and the polarization angle $\varphi_0 = \pi/4$. We emphasize that both beams have no circular polarization. Yet, they induce spatially textured spin currents giving rise 
	to a spin separation with a quite significant local spin density. The band-structure and kinetic parameters used in the calculations are relevant to transition metal dichalcogenide (TMDC) layers, they are given in the figure caption. The used spin relaxation time $\tau_s = 10$~ns 
	is consistent with experiments on electron spin dynamics in TMDC monolayers~\cite{Yang:2015, Dey:2017}.
	
	Despite different patterns of the spin density, cf. Figs.~\ref{Figure1}(c) and~\ref{Figure1}(f), the dominant contribution to the spin separation in both cases 
	originates from spatial variation of the Stokes parameters $S_1$ and $S_2$ in the beams. We recall that $S_1 = |E_x|^2 - |E_y|^2$ and $S_2 = (E_x E_y^* + E_y E_x^*)$ define the local linear polarization in the $(x,y)$ axes and the diagonal axes, respectively. The corresponding spin current in Eq.~\eqref{js} is proportional to the parameter $\mathcal W_2$. For the considered beams, its radial and azimuthal components are given by $j_{s, r} \propto 2 e_r e_\varphi$ and $j_{s, \phi} \propto e_r^2 - e_\varphi^2$, where $e_r$ and $e_\varphi$ are the components of the local polarization vector in the polar coordinate frame. The spin separation is determined by $\mathrm{div}\bm j_{s} = \mathcal W_2 [2 \partial^2_{xy} S_1 - (\partial_{x^2}^2 - \partial_{y^2}^2) S_2]$.

	For vector beams in the form of Eq.~\eqref{En},  only the polarization mechanism contributes to spin separation. It gives radially symmetric distributions $\bm j_s(\bm r)$ and $s(\bm r)$ for the beam with the winding number $n=1$, Figs.~\ref{Figure1}(e) and~\ref{Figure1}(f). The amplitude (and the sign) of $s(\bm r)$ follows $\sin(2 \varphi_0)$
	and is, therefore, controlled by the polarization angle $\varphi_0$ (see Appendix). 
	Particularly, radially or azimuthally polarized vector beams (with $\varphi_0 = 0$ or $\varphi_0 = \pi/2$, respectively) generate pure vortex spin currents with $\mathrm{div}\bm j_{s} = 0$  resulting in no spin separation. 
	The spin separation is most efficient for the polarization angles $\varphi_0 = \pm \pi/4$, i.e., the field $\bm E \parallel (\bm e_r \pm \bm e_\varphi)/\sqrt{2}$,
	see Fig.~\ref{Figure1}(f).
	
	For twisted beams in the form of Eq.~\eqref{El}, the polarization mechanism gives the spin distribution $s(\bm r) \propto \sin(2\varphi - 2 \varphi_0)$ whose radial structure depends on $|l|$ (see Appendix). This second-angular-harmonic contribution prevails in the spin pattern in Fig.~\ref{Figure1}(c). Additionally, there are contributions stemming from the spin currents $\propto \mathcal W_4$ and $\propto \mathcal W_5$ in Eq.~\eqref{js}. These currents, originating from the 
	rotational photon drag together with the spin Hall effect, are sensitive to the phase profile of the electromagnetic field and are odd in the OAM projection $l$.  
	The current $\propto \mathcal W_4$ is insensitive to the field polarization and has only radial  component, whereas  the current $\propto \mathcal W_5$ is sensitive to the linear polarization and has both radial and azimuthal components. 
	The current $\propto \mathcal W_4$  results in a radially symmetric profile $s(\bm r)$ with a finite spin density at the beam center controlled by the OAM. This non-zero spin density at the beam center is clearly visible in Fig.~\ref{Figure1}(c).
	
	Generally, the emergent spin density in the 2D plane can be represented in terms of the angular harmonics 
	\,\\[-2.8ex]\begin{multline}
		\label{s_m}
		s(r,\varphi) =  \sum_{m \geqslant 0} c_m(qr) \cos (m\varphi ) +
		\sum_{m \geqslant 1} s_m(qr) \sin (m\varphi ) \,,
	\end{multline}\,\\[-5ex]
	where $c_m(x)$ and $s_m(x)$ are the radial functions. Our analysis in Appendix shows that $s(r,\varphi)$ contains only the zero and second angular harmonics for twisted beams in the form of Eq.~\eqref{El} and the harmonics $|2(n-1)|$ for the vector beams in the form of Eq.~\eqref{En}.
	
\begin{figure}[t]
		\includegraphics[width=0.95\linewidth]{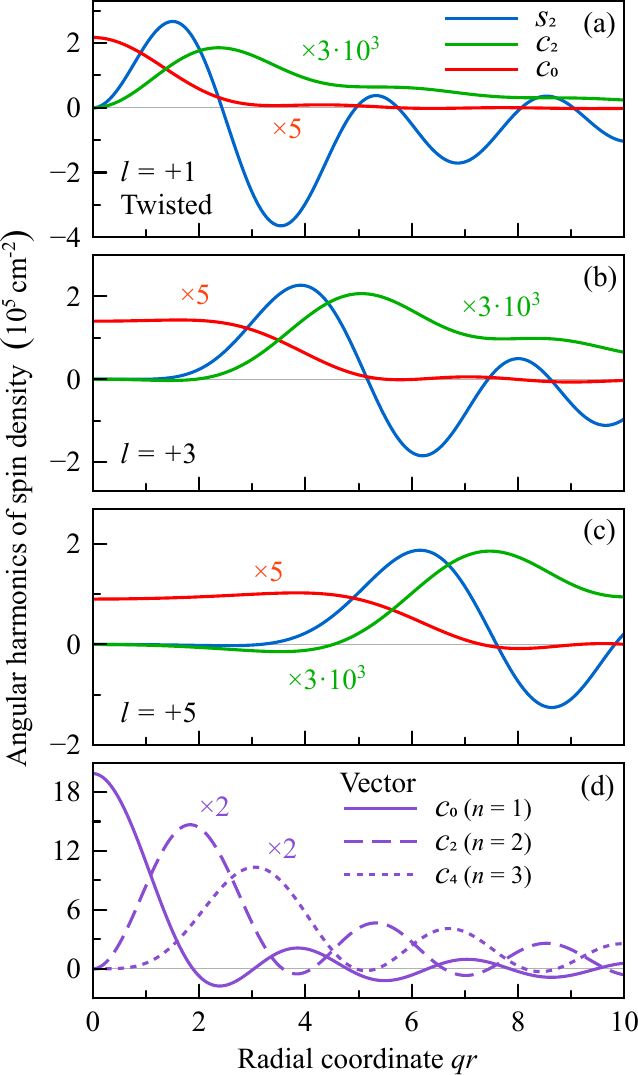}
		\caption{Radial functions $c_m(qr)$ and $s_m(qr)$ of the spin density expanded over angular harmonics, see Eq.~\eqref{s_m}.
               (a)-(c) Evolution of spin distributions induced by twisted Bessel beams with increasing the OAM projection $l$. 
               (d) Spin distributions induced by vector beams with different winding numbers $n$. 
               The distributions are calculated for the same parameters as in Fig.~\ref{Figure1}, relevant to $n$-type TMDC layers, for 
               polarization angles (a)-(c) $\varphi_0 = 0$ and (d) $\varphi_0 = \pi/4$.}
		\label{Fig2}
	\end{figure}

	Figures~\ref{Fig2}(a)-(c) show how the radial functions $c_0(qr)$, $c_2(qr)$, and $s_2(qr)$ evolve with the increase of the OAM for twisted beams. These functions are proportional to the parameters $\mathcal W_4$, $\mathcal W_5$, and $\mathcal W_2$, respectively. Typically, the parameters 
	$\mathcal W_{4,5}$ are smaller than $\mathcal W_2$, since the corresponding spin currents arise from the conversion of charge photon drag currents  via the spin Hall effect, see Eq.~\eqref{Wj}. Also, for the near band-edge excitation, i.e. for $\hbar \omega - 2\delta \ll 2\delta$, the $\mathcal W_5$ parameter contains additional small factor $(w^2-1)^2$ resulting in $\mathcal W_5 \ll \mathcal W_4$. Figures~\ref{Fig2}(a)-(c) reveal that the increase of the OAM leads to the growth of the radii of central rings in the spatial distributions of $s_2(qr)$ and $c_2(qr)$, reflecting the spatial profile of the Bessel beam intensity proportional to $J_l^2(qr)$. On the other hand, the radially symmetric contribution $c_0(qr)$ remains finite inside the first ring, despite the local radiation intensity is nearly zero. This is the result of spin diffusion, which is quite efficient for the chosen parameters with $L \approx 5.6~\mu$m and $l_s \approx 3.7~\mu$m. The total spin accumulated in this region 
	increases with the OAM, and its sign is determined by the sign of $l$.

	Figure~\ref{Fig2}(d) shows the radial functions for the spin density induced by vector beams with various winding numbers $n$. We fix the polarization 
	angle at $\varphi_0 = \pi/4$. In this case, only the functions $c_{2n-2}(qr)$ are present in the spin distribution, see Appendix; these functions are plotted in Fig.~\ref{Fig2}(d).
        For the winding number $n=1$, the angle between the local polarization vector $\bm e$ and the position vector $\bm r$ is fixed (and equal to $\varphi_0$).
        Therefore, the spin density is rotationally symmetric and finite at the beam center. 
        For $n \neq 1$, the direction of $\bm e$ with respect to $\bm r$ rotates with the polar angle $\varphi$, resulting in vanishing spin polarization at the beam center.
        The maximum spin polarization is reached at $qr \sim (n-1)$. 
        Figure~\ref{Fig2}(d) shows the radial functions for positive winding numbers $n$. Interestingly, for $n \leqslant 0$, the spin distribution coincides with that for the 
        winding number $2-n$ despite the different intensity and polarization patterns of the beams, see Appendix.

	\textit{Conclusions.} 
	To summarize, we have demonstrated that twisted and vector optical beams with local linear polarization can spatially separate 2D Dirac electrons with opposite spin projections. The mechanism is based on the light-structure-controlled generation of diverging spin currents, which result in spin accumulation.
	We have derived analytical equations for the resulting spin density in the quasi-local and diffusion-controlled regimes, distinguished by the ratio of the spin diffusion length to the characteristic radius of the beam intensity rings. In the intermediate regime, we have numerically calculated the emerging spin textures and analyzed their dependence on the beam OAM and polarization winding number for TMDC monolayers. 
	The spatial distribution of the spin density can be experimentally probed using Faraday or Kerr rotation techniques. 
	
	\textit{Acknowledgments.}  This work was supported by the Russian Science Foundation (Project No. 22-12-00211-$\Pi$). A.A.G. and M.V.D. also acknowledge the support from the Basis Foundation for the
Advancement of Theoretical Physics and Mathematics.

	\textit{Data availability.} The data that support the findings of this study are available from the corresponding author upon reasonable request.
	
	\setcounter{equation}{0}
	\renewcommand{\theequation}{A\arabic{equation}}
	
	\textit{Appendix: Calculation of spin density.} Here, we derive analytical equations for the spin density $s(\bm r)$ in the limiting cases of the quasi-local and diffusion-controlled regimes and
	provide details of numeric calculations of $s(\bm r)$ at arbitrary $l_s/L$.
	
	First, we use Eq.~\eqref{js} to calculate the spin currents $\bm j_{s}$ and the generation term $\mathrm{div}\bm j_{s}$. Then, we derive the spin density in the quasi-local approximation $s^{\mathrm{(loc)}}(\bm r)$ using Eq.~\eqref{s_local}. For calculation of the spin distribution $s^{\mathrm{(diff)}}(\bm r)$ in the diffusion-controlled regime after Eq.~\eqref{s_diff}, we use the decomposition (known from electrostatics)
		\begin{equation}\label{log_expansion}
			\mathrm{ln}\,|\bm x-\bm x'| = \mathrm{ln}\,x_> - \sum_{m\ne0}\frac{1}{2|m|}\left(\frac{x_<}{x_>}\right)^{\!|m|}e^{\i m(\varphi-\varphi')} \,,
		\end{equation}
		where $\bm x$ and $\bm x'$ are two-dimensional vectors, $x_<$ and $x_>$ are $\mathrm{min}(x,x')$ and $\mathrm{max}(x,x')$, respectively, and $\varphi$ and $\varphi'$ are the polar angles of the vectors $\bm x$ and $\bm x'$, respectively. Equation~\eqref{log_expansion} allows us to integrate Eq.~\eqref{s_diff} over the polar angle $\varphi'$ and obtain the spin distribution in the polar coordinate frame. The results are summarized below.

	(i) For twisted Bessel beams of the form Eq.~\eqref{El}, the spin currents with nonzero divergence are given by
		\begin{eqnarray}
			j_{s,r} = E_0^2 \bigl\{ qJ_l(J_{l-1} - J_{l+1}) \mathcal W_2 \sin(2\varphi- 2\varphi_0) \nonumber  \\
			- \frac{l}{r}J^2_l \left[ \mathcal W_4 - \mathcal W_5 \cos(2\varphi-2\varphi_0)\right] \bigr\}\:, \nonumber \\[1ex]
			j_{s,\varphi} = E_0^2 \bigl\{qJ_l(J_{l-1} - J_{l+1}) \mathcal W_2\cos(2\varphi- 2\varphi_0) \nonumber  \\
			- \frac{l}{r}J^2_l \mathcal W_5 \sin(2\varphi-2\varphi_0) \bigr\}\:,
		\end{eqnarray}
		where $J_l$ is a contraction for $J_l(qr)$.
	
		In the quasi-local approximation, the spin density has the form
		\begin{multline}
			\label{sloc_twisted}
			s^{\mathrm{(loc)}}(r,\varphi) = -\tau_s E_0^2 \biggl[ q \left( q f_1'-\frac{f_1}{r} \right) \mathcal W_2 \sin (2\varphi- 2\varphi_0) \\
			- \frac{q l}{r}f_2'\,\mathcal W_4 + l \left( \frac{q}{r}f_2'-\frac{2 f_2 }{r^2} \right) \mathcal W_5 \cos (2\varphi- 2\varphi_0) \biggr] \:,
		\end{multline}
		where 
		\begin{equation}
			f_1 = J_l(qr)\bigl[J_{l-1}(qr) - J_{l+1}(qr)\bigr]\:,~f_2 = J_l^2(qr)\:.
		\end{equation}

		In the diffusion-controlled regime, one obtains
		\begin{multline}
			s^{\mathrm{(diff)}}(r,\varphi) = \frac{E_0^2}{2D_s}\biggl[2 J_{l-1}J_{l+1}  \mathcal W_2  \sin\left(2\varphi - 2\varphi_0\right) \\
			-  \frac{l}{|l|}\left(J_l^2 - \sum\nolimits_{k=-|l|}^{|l|}J^2_{k}\right) \mathcal W_4\\ + l \bigl(J_l^2 - J_{l-1}J_{l+1}\bigr) \mathcal W_5 \cos\left(2 \varphi - 2\varphi_0 \right) \biggr]\:.
		\end{multline}

	(ii) For vector Bessel beams of the form Eq.~\eqref{En}, the currents are given by
		\begin{eqnarray}
			j_{s,r} &=& -2  E_0^2 q J_nJ_{n-1} \, \mathcal W_2 \sin\left[2(n-1)\varphi + 2\varphi_0\right]\:, \nonumber \\[1ex]
			j_{s,\varphi} &=& 2 E_0^2 q  J_nJ_{n-1} \, \mathcal W_2 \cos\left[2(n-1)\varphi + 2\varphi_0\right]\:.
		\end{eqnarray}
		The spin density in the quasi-local approximation and the diffusion-controlled regime has the form
		\begin{multline}
			\label{sloc_vector}
			s^{\mathrm{(loc)}}(r,\varphi) = 2\tau_s E_0^2q^2 \bigl(J_{n-1}^2 + J_{n}J_{n-2}\bigr) \\ \times \mathcal W_2 \sin\bigl[2(n-1)\varphi + 2\varphi_0\bigr]\:
		\end{multline}
		and
		\begin{equation}
			s^{\mathrm{(diff)}}(r,\varphi) = \frac{E_0^2}{D_s} J_{n-1}^2 \mathcal W_2 \sin\bigl[2(n-1)\varphi + 2\varphi_0\bigr] \,,
		\end{equation}
respectively.
	
For numeric calculation of $s(\bm r)$ in the general case of arbitrary relation between $l_s$ and $L$, we follow Eq.~\eqref{general solution} and use the decomposition
		\begin{equation}\label{K0_expansion}
			K_0(|\bm x-\bm x'|) = \sum_m I_m(x_<)K_m(x_>)\,e^{\i m(\varphi-\varphi')} \,,
		\end{equation}
		where $I_m(x)$ and $K_m(x)$ are the modified Bessel functions. 
		Expanding also the generation term $\mathrm{div}\bm j_{s}(\bm r)$ via the angular harmonics as follows
		\begin{equation}
			\mathrm{div}\bm j_{s} =  \sum_{m \geqslant 0} a_m(qr) \cos (m\varphi ) 
			+
			\sum_{m \geqslant 1} b_m(qr) \sin (m\varphi ) \,,
		\end{equation}
		we obtain Eq.~\eqref{s_m} with
		\begin{align}
			\label{cm_sm}
			c_m =& -\frac{1}{D_s} \int a_m(qr')  I_m\left(\frac{r_<}{l_s}\right)K_m \left(\frac{r_>}{l_s} \right)\,r'\mathrm{d} r'\:, \nonumber\\
			s_m =& -\frac{1}{D_s} \int b_m(qr')  I_m\left(\frac{r_<}{l_s}\right)K_m \left(\frac{r_>}{l_s} \right)\,r'\mathrm{d} r'\:.
		\end{align} 
The angular harmonics $a_m$ and $b_m$ are readily found from Eqs.~\eqref{sloc_twisted} and \eqref{sloc_vector} since $\mathrm{div}\bm j_{s} = -s^{\rm (loc)}/\tau_s$.  Then, Eq.~\eqref{cm_sm} is used to calculate the spin density distributions plotted in Fig.~\ref{Figure1} and Fig.~\ref{Fig2}.

	\bibliographystyle{apsrev4-1-customized}
	\bibliography{bibliography}

\end{document}